\documentstyle[aps,prb,multicol,epsf,graphicx]{revtex}

\newcommand \be {\begin{equation}}
\newcommand \ee {\end{equation}}
\newcommand \eps {\epsilon}
\newcommand \bi {\bibitem}

\begin{document}

\title{HEAT TRANSFER AND FOURIER'S LAW IN OFF-EQUILIBRIUM SYSTEMS}
\author{A.Garriga and F.Ritort}
\address{Departament de F\'{\i}sica Fonamental, Facultat de F\'{\i}sica, Universitat de 
Barcelona\\
Diagonal 647, 08028 Barcelona (Spain)\\
E-Mail: adan@ffn.ub.es, ritort@ffn.ub.es}

\date{\today}

\maketitle
\begin{abstract}                                        
We study the most suitable procedure to measure the effective
temperature in off-equilibrium systems. We analyze the stationary
current established between an off-equilibrium system and a thermometer
and the necessary conditions for that current to vanish. We find that
the thermometer must have a short characteristic time-scale compared to
the typical decorrelation time of the glassy system to correctly measure
the effective temperature. This general conclusion is confirmed
analyzing an ensemble of harmonic oscillators with Monte Carlo dynamics
as an illustrative example of a solvable model of a glass. We also find
that the current defined allows to extend Fourier's law to the
off-equilibrium regime by consistently defining effective transport
coefficients. Our results for the oscillator model explain why thermal
conductivities between thermalized and frozen degrees of freedom in
structural glasses are extremely small.
\end{abstract}       

\bigskip
\section{INTRODUCTION}
In the last decades there has been an increasing interest in trying to extend the thermodynamics ideas to systems which are far out of
equilibrium. The general problem is to quantify non-equilibrium fluctuations
within a thermodynamic approach. Recent studies suggest that the slow dynamics of glassy systems allows us to do such kind of generalizations\cite{SIT}. In this direction, it has been proposed a definition of a
non-equilibrium temperature (also called {\it effective temperature})
based on the violation of the fluctuation-dissipation theorem
(FDT)\cite{CUG,FRA,BOU,KUR}. This temperature verifies the relation:

\begin{equation}
T_{\rm{eff}}(t',t)R(t',t)= \frac{\partial C(t',t)}{\partial t'}~~~~~~t~>~t'~,
\label{FDT}
\end{equation}
where $ C(t',t) $ is the autocorrelation function and $ R(t',t) $  is the corresponding response function. It has been suggested that for a large class of glassy systems in the long-time regime ($ t' \rightarrow \infty $) the effective temperature is only a function of the smallest time, i.e $T_{\rm{eff}}(t',t)\rightarrow T_{\rm{eff}}(t') $ and also that is a good quantity in the thermodynamic sense \cite{KUR,KUR2,NIEW2}. This knowledge is mainly theoretical and comes from the analysis of mean-field glassy systems. There is also good numerical support to these assumptions for structural glass models\cite{NUM}, although from the experimental point of view, the question is still far from settled (see, for example, recent experimental results on glycerol\cite{EXP} and laponite\cite{cil}). Note that in the general framework of thermodynamics, the concept of temperature is associated with local equilibrium, i.e the fact that local regions in space admit a thermodynamic description in terms of space-dependent intensive variables such as the temperature.

In this paper we try to understand the most adequate procedure to measure the effective temperature. This question has been recently addressed by Cugliandolo, Kurchan and Peliti\cite{KUR} and Exartier and Peliti\cite{EXH} who have discussed possible procedures to measure the effective temperature.

In a previous paper we have used the concept of the effective temperature in order to investigate the validity of the zero-thermodynamic law in off-equilibrium systems\cite{art1}. We have seen that in absence of coupling via the dynamics two coupled systems far from equilibrium never thermalize. This poses the question: why different degrees of freedom in structural glasses can stay at different effective temperatures forever?. In this paper we shall show that the answer comes from the smallness of the conductivity between the two systems in the aging state.

The paper is organized as follows. In section II we define the heat transfer between two generic coupled systems. In section III we analyze the heat transfer between an off-equilibrium system and a thermometer in the framework of Linear Response Theory (LRT). In section IV we show an explicit example. In section V we extend Fourier's law to the off-equilibrium regime. In section VI there are the conclusions.

\section{DEFINITION OF THE HEAT TRANSFER}
In systems which are in equilibrium, the concept of a thermometer is related to the zeroth law of thermodynamics. If we want to measure the temperature of a system in equilibrium with a thermal bath at temperature $ T $, we take another macroscopic system (called thermometer) in contact with its own thermal bath. Then we put the thermometer in contact with the system in such a way that the system is weakly perturbed. If the temperature of the thermometer's bath is different from $ T $ , the whole system (thermometer + system) is off-equilibrium, as a consequence there is a heat transfer from the hot system to the cold one. In fact, in order to measure $ T $  we change the temperature of the thermometer's thermal bath until we cannot measure any heat transfer between the system and the thermometer; then, the whole system is in equilibrium and the temperature of thermometer's bath is just $ T $ .\\
For the non-equilibrium case we shall follow the same steps. We couple the off-equilibrium system with a thermometer which is in equilibrium with a thermal bath at a certain temperature. If perturbed, this thermometer relaxes to equilibrium with a characteristic time-scale (in this case the perturbation is the coupling with the off-equilibrium system). In order to have some kind of ``off-equilibrium zeroth law'' we have to define the heat transfer between the two systems. Let us suppose that we have two generic systems, $ X $  and $ Y $ with $ N $  internal degrees of freedom $ x_i $  and $ y_i $  respectively and we couple them. Assuming that the coupling between the systems is small, we may write for the total energy: 
\begin{equation}
H \longrightarrow H - \epsilon \sum_{i=1}^N x_i y_i~~.    
\label{ham} 
\end{equation} 
We consider the two-time correlation and overlap functions defined in the following way:
\begin{eqnarray}
 C_X(t,t') =\frac{1}{N} \sum_{i=1}^N x_i(t)  x_i(t');~~~                 
 C_Y(t,t') =\frac{1}{N} \sum_{i=1}^N y_i(t)  y_i(t')\\
 Q_1(t,t') =\frac{1}{N} \sum_{i=1}^N y_i(t)  x_i(t');~~~
 Q_2(t,t') =\frac{1}{N} \sum_{i=1}^N x_i(t)  y_i(t')~.
\label{overl} 
\end{eqnarray}
Following \cite{KUR,EXH} we define the heat transfer between them as the net balance of the power supplied by one system to the other. In terms of the overlap functions the heat transfer $ J(t,t') $ reads:
\begin{equation}
J(t,t') =\eps \frac{\partial}{\partial t} (Q_2(t,t') - Q_1(t',t))=  \epsilon \sum_{i=1}^N \left( \dot{x}_i(t)y_i(t') - x_i(t')\dot{y}_i(t) \right).
\label{curr}
\end{equation}
The heat transfer vanishes in equilibrium because of the time-reversal symmetry property of correlation functions, i.e $Q_1(t,t') = Q_2(t,t') = Q(|t-t'|) $. If there is a system which is out of equilibrium this quantity is, in general, non-zero. From now on we shall assume that the system $ X $  is the thermometer, while $ Y $  will be the off-equilibrium system which is relaxing towards its equilibrium state. Depending on the ratio between the characteristic time-scale of the thermometer and the relaxation time-scale of the off-equilibrium system we can identify the different contributions of each term in the right hand side (r.h.s.) of (\ref{curr}): 
\begin{itemize} 
\item If the relaxation time of the thermometer is fast enough, only the second term in the r.h.s. of (\ref{curr}) contributes. The contribution of the first term is negligible due to the fact that the net force made by the fast degrees of freedom of the thermometer upon the system is zero on average. Note that in\cite{KUR} the definition of heat transfer contains only the second term in the r.h.s of (\ref{curr}); although not explicitly stated there, in this definition it is implicitly assumed the fact that the characteristic time-scale of the thermometer is much smaller than the relaxing time-scale of the off-equilibrium system. \item If the thermometer has a relaxation time larger than the characteristic time of the system, the main contribution to $ J(t,t') $  comes from the first term in the r.h.s of (\ref{curr}). We will see that in this case we don't have a good thermometer for measuring effective temperatures. Obviously, when the typical time of the thermometer is of the order of the relaxation time of the system, the two terms in the r.h.s of (\ref{curr}) are of the same order, and we must to take into account both of them.
\end{itemize}
It is believed that for measuring the effective temperature the characteristic time-scale of the thermometer should be of the same order as the relaxation time-scale of the system. This is not clear at all because the power supplied by the thermometer to the system must be negligible in order to measure correctly the effective temperature, so the term $ \dot{x}(t) y(t') $ should be smaller enough in comparison to the term  $ x(t') \dot{y}(t) .$ This condition can be fulfilled only in the case of faster thermometers. Moreover, if the time-scale of the thermometer is longer than the relaxation time-scale of the system, the two systems are off-equilibrium during the measurement due to the fact that the perturbed thermometer has not time enough to relax to its equilibrium state. In fact, the temperature of the thermometer is only well defined when it is in equilibrium.
\section{LINEAR RESPONSE THEORY}

Now suppose that  $ \epsilon $ is small enough so that we can consider
the $\epsilon $-coupling term in (\ref{ham}) as a perturbation. Thus the
effective temperature will be slightly perturbed. Hence, using linear
response theory (hereafter referred to as LRT), we have:   
\begin{equation}
x_i^{\epsilon}(t) = x_i^{\epsilon = 0}(t) + \epsilon \int_s^t du R_X(t,u)y_i(u);~~~~
y_i^{\epsilon}(t') =  y_i^{\epsilon = 0}(t') + \epsilon \int_s^{t'} du R_Y(t',u)x_i(u)~. 
\label{LRT}
\end{equation}
Here we have considered that the two systems are coupled at a time $t=s$ (the ``coupling time'');  so, we must integrate the responses of the system and the thermometer between two times: the coupling time $s$ and the time we want to measure. 
For the general non-equilibrium case, if we assume that $ T_{\rm{eff}}(t',t) $ only depends on the smallest time, we obtain in the limit when the two times involved are much bigger than the coupling time $s$: 
\begin{eqnarray}
\lim_{t'\rightarrow t} J = \epsilon^2 \left[  \left(\frac{\partial C_Y(t,u)}{\partial u}\right)_{u=t} C_X(t,t)(\beta_X - \beta_Y(t))\right]
+  \epsilon^2 \left[ \int_s^t du R_X(t,u) \left[ -\frac{\beta_Y(u)}{\beta_X} \frac{\partial C_Y(t,u)}{\partial u} - \frac{\partial C_Y(t,u)}{\partial t} \right] \right] \nonumber\\
- \epsilon^2 \left[  \int_s^t du C_X(t,u) \left[ \beta_X \frac{\partial^2 C_Y(t,u)}{\partial u^2} + \beta_Y(u) \frac{\partial^2 C_Y(t,u)}{\partial u \partial t} \right]\right]~,
\label{current}
\end{eqnarray}
where we have used (\ref{LRT}) and the generalization of FDT (\ref{FDT}), being $ \beta_Y(u)=1/T_{\rm{eff}}(u) $ and $ \beta_X=1/T_X $. Is easy to verify that the expression for the current in the equilibrium case reduces to (\ref{current}) with $ \beta_Y(u)= \beta_Y $, yielding:
\begin{equation}
\lim_{t \rightarrow t'} J(t,t') =  2\epsilon^2 (\beta_X -\beta_Y)\int_{s}^t du \frac{\partial C_X(t-u)}{\partial t} \frac{\partial C_Y(t-u)}{\partial t},
\label{currenteq}
\end{equation} 
which is some kind of Green-Kubo relation \cite{KUB}. From this expression we realize immediately that in the equilibrium case  the heat transfer is zero only when the two systems are at the same temperature, and it remains equal to zero for all times.
At this point we have to explain better the measurement procedure. As the off-equilibrium system is relaxing towards its equilibrium state, we cannot find a single temperature thermometer that cancels the heat transfer during all the time. This is due to the assumption that the thermometer is a system in equilibrium while the effective temperature of the off-equilibrium system depends on time. In order to measure the effective temperature at a time $ t $ we have to couple the thermometer at a time $ s $. We shall see later that we can take $ s=t-n\tau $ where $ \tau $ is the characteristic time-scale of the thermometer and $ n $ a large integer number. Physically the result is independent of $ n $ provided that $ n >> 1 $ .  Obviously, for every time $t$  we are interested in, the measured temperature will be different as we expect. It is easy to prove that only when the characteristic time of the thermometer $ \tau $ is much less than the relaxing time of the system, i.e $ \tau << t $ , we have:
\begin{equation}
\lim_{t\rightarrow \infty} \frac{(\beta_X-\beta_Y(t))}{\beta_Y(t)} \longrightarrow 0~~, 
\label{beta1}
\end{equation}
otherwise, for the limit $ \tau >> t $, we have:
\begin{equation}
\lim_{t\rightarrow \infty} \frac{(\beta_X-\beta_Y(t))}{\beta_Y(t)} \neq 0~~,
\label{beta2}
\end{equation}
and the thermometer cannot properly measure the effective temperature.

\section{an example}

As an instructive example we consider a model of uncoupled harmonic oscillators with parallel Monte Carlo dynamics introduced in~\cite{BPR} and recently revisited in \cite{art1,TEO,CRI}. This is a simple model which contains the essential features of glasses such as aging in the correlation and response functions. This model is defined by the Hamiltonian:
\begin{equation}
H = \frac{1}{2} K \sum_{i=1}^N x_i^2~~.
\end{equation}
 The Monte Carlo dynamics is implemented by small movements: $ x_i \rightarrow x_i + r_i/\sqrt{N} $  where $ r_i $  are random variables Gaussian distributed with zero average and variance $ \Delta^2. $ The move is accepted according to the Metropolis algorithm with probability $ W(\Delta E) $ which satisfies detailed balance: $  W(\Delta E) =  W(-\Delta E) \exp (-\beta \Delta E ), $ where $ \Delta E $  is the change in the Hamiltonian.\\
 Here we will concentrate our attention on the dynamics at zero
 temperature which is known to be glassy. The interest of the
 zero-temperature dynamics relies on the extremely slow decay of the
 energy which decays like $ 1/\log (t) $ due to the entropic barriers
 generated by the low acceptance rate. For our purpose, we need the
 correlation function defined as $ C_Y(t,t') =\frac{1}{N} \sum_i^N
 y_i(t)  y_i(t')  $ and the response function: $ R_Y(t,t') = \left(
 \frac{\delta M(t)}{\delta h(t')} \right)_{h=0}$ with $t>t'$,   
where the magnetization for this model is $ M(t)=\frac{1}{N} \sum_i^N y_i(t). $

The results are \cite{BPR}:
\begin{equation}
C_Y(t,t') = \frac{2E_Y(t')}{K} \exp\left( -\int_{t'}^t f_Y(x) dx \right);~~~~~~
R_Y(t,t') = \frac{f_Y(t')}{K} \exp\left( -\int_{t'}^t f_Y(x) dx \right) \Theta (t-t')~~,
\label{BPRresp} 
\end{equation}
where $ f_Y(t) $ is a function that contains the whole dynamics and decays like $ 1/t $ at zero temperature\cite{BPR}. The effective temperature (\ref{FDT}) can be exactly computed for this model:
\begin{equation}
T_{\rm{eff}}(t') = 2E_Y(t') + \frac{2}{f_Y(t')} \frac{\partial E_Y(t')}{\partial t'}~~.
\label{teffosc}
\end{equation}
Note that for this particular case, the effective temperature is only a function of the smallest time. This is a characteristic feature of this particular model that is believed to hold for structural glass models in the asymptotic limit $ t' \rightarrow \infty .$ In this limit (\ref{teffosc}) verifies $T_{\rm{eff}}(t')\rightarrow 2E_Y(t') $ at zero temperature.

To measure this effective temperature we have to define a general thermometer (i.e a system in equilibrium which verifies FDT). In general we can write for the correlation and the response functions of the thermometer:
\begin{equation}
C_X(t-t') = C_X(0)e^{-(t-t')/\tau}; ~~~~~
R_X(t-t') = \frac{\beta_X}{\tau} C_X(t-t')~~.
\end{equation}
Notice that if the characteristic time of the thermometer $ \tau $  is very small (a fast thermometer) the correlation function $ C_X(t-u) $ only contributes if $ u \approx t $ ; in this case we can write $ \frac{\partial C_Y(t,u)}{\partial t} = - \frac{\partial C_Y(t,u)}{\partial u}.$ This result is valid in the $ t \rightarrow \infty$ limit and comes from the fact that, at very short time-scales, the aging system can be considered as if it were in equilibrium (at the corresponding effective temperature) and the partial derivatives of the correlation functions can be interchanged in the usual way. This can be checked by explicit derivation of eq.(\ref{BPRresp}). Then, the heat transfer will vanish if the condition $ \beta_Y(t) = \beta_X $ is satisfied. Therefore, if we have a very fast thermometer we can measure exactly the effective temperature. By numerical evaluation of the integrals (\ref{current}) we compute the temperature $ T_m $  of the thermometer which makes the heat transfer vanish. The results for different thermometers are shown in the left panel of Fig.1 where one can clearly see that for $ t >> \tau $ the measured temperature $ T_m=1/\beta_m $ is always higher than the effective one and converges to it logarithmically. On the contrary, for  $ t << \tau $ the thermometer does not measure (for a given time $t$) the right effective temperature. We should say that for $ t << \tau $ the integrals in (\ref{current}) were evaluated taking $s=0$. In the right panel of Fig.1 we can also see the behavior of the measured temperature that shows some qualitative agreement with experimental results \cite{EXP}.

Now, as we know the asymptotic behavior of the correlation and response, we can compute explicitly the large time evolution of the difference between the effective temperature and the measured one. A saddle-point calculation of (\ref{current}) shows that for $ t \longrightarrow \infty$:
\begin{equation}
\frac{\beta_m-\beta_Y(t)}{\beta_Y(t)}=\frac{\frac{\partial E_Y(t)}{\partial t}}{2E_Y(t)f_Y(t)}=-\frac{1}{2} \frac{1}{\log\left( \frac{2t}{\sqrt{\pi}} \right)}\left( 1- \frac{1}{2} \frac{\log(\log\left( \frac{2t}{\sqrt{\pi}} \right))}{\log\left( \frac{2t}{\sqrt{\pi}} \right)} \right)~~.
\label{asimpt}
\end{equation}
In this calculation we made the assumption that due to the smallness of the relaxation time of the thermometer, the main contribution in the integral term of (\ref{current}) is just when $ u = t .$ An important remark is that in this limit the difference does not depend on $ \tau $ provided that it is small enough ($\tau/t << 1)$. This is clearly seen in the inset of the left panel in Fig.1. If we consider only the second term in the r.h.s. of (\ref{curr}) (as proposed in \cite{KUR}) we also find that the measured temperature converges to the effective one but the logarithmic asymptotic behavior in (\ref{asimpt}) is different. In fact the result for this case is $\frac{\beta_m-\beta_Y(t)}{\beta_Y(t)}=\frac{\frac{\partial E_Y(t)}{\partial t}}{E_Y(t)f_Y(t)} .$ So, in order to obtain the correct logarithmic corrections to the asymptotic long-time behavior ( $\frac{\beta_m-\beta_Y(t)}{\beta_Y(t)} \rightarrow 0)$  we must consider the two terms in the expression for the current (\ref{curr}).

\begin{figure}[tbp]
\begin{center}
\includegraphics*[width=16cm,height=8cm]{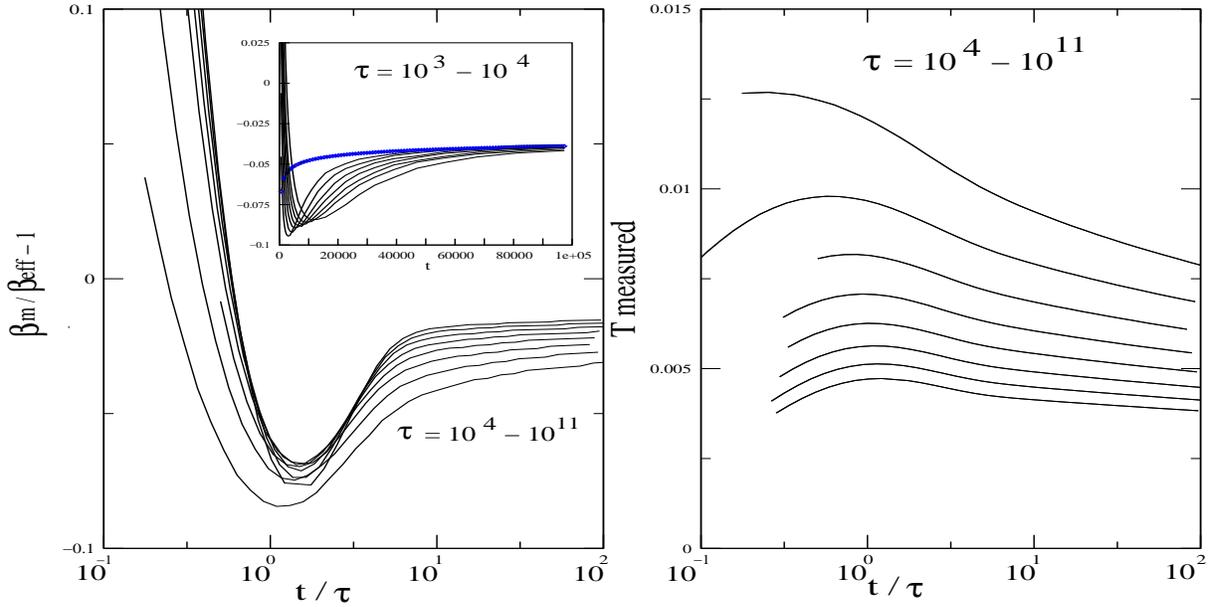}
\vskip 0.1in
\caption{On the left there is a plot of the difference between the measured temperature and the effective one in function of the rescaled time (in Monte Carlo step unities) of measure. The thermometers have characteristic time-scales in the range: $  10^4 - 10^{11} $. The lowest curve corresponds to the shorter time-scale and the highest to the largest one. In the inset we have shown the asymptotic behavior for thermometers in the range: $\tau= 10^3 - 10^4 $ being the upper line the analytical result (\ref{asimpt}). On the right there is a plot of the measured temperature as function of the rescaled time for the same thermometers. The highest curve corresponds to the quickest thermometer.  
\label{fig1}}
\end{center}
\end{figure}


\section{FOURIER'S LAW}

We have seen that the off-equilibrium current we have defined is a quantity which can be used to measure the effective temperature for systems which are off-equilibrium. Now we can think about the heat transfer itself. In standard non-equilibrium thermodynamics there is a linear relationship between the temperature gradient and the subsequent current established\cite{MAZUR}. This is the well-known Fourier's law: 
\be
J=L_{QQ} \nabla (\frac{1}{T}) = -\kappa \nabla T~~,
\ee 
where $ L_{QQ} $ is the Onsager coefficient and $ \kappa $ is the thermal conductivity. For the equilibrium case we can compute the heat transfer (\ref{currenteq}) using (\ref{BPRresp}):
\begin{equation}
\lim_{t'\rightarrow t} J(t,t') = 4 \epsilon^2 \frac{C_X(0)f_Y E_Y}{K_Y} (\beta_X -\beta_Y)~~,
\end{equation}
Using the asymptotic expansions for $ f_Y(t) $ and $E_Y(t) $ (for a recent review see \cite{art1}) one can see that in the long-time limit, the non-equilibrium case satisfies Fourier's law: 
\begin{equation}
\lim_{t'\rightarrow t} J(t,t') \approx  \epsilon^2 \frac{1}{\left( \log \left( \frac{2t}{\sqrt{\pi}} \right) + \frac{1}{2} \log(\log\left( \frac{2t}{\sqrt{\pi}} \right) ) \right)^{ \frac{1}{2}}}  \frac{1}{t \log^{\frac{1}{2}} \left( \frac{2t}{\sqrt{\pi}} \right)} (\beta_X -\beta_Y(t))~~.
\end{equation}
So the Onsager coefficient decays as  $ L_{QQ} \simeq 1/(t\log(t)).$ As in the long-time limit the effective temperature ($\beta_Y(t))$ decays like the energy, the thermal conductivity decays as $ \kappa \simeq \log(t)/t.$ This simple example extends the Fourier's law to aging systems. So this is an extension of the linear relations between fluxes and forces obtained in the framework of the non-equilibrium thermodynamics \cite{MAZUR}, but with time-dependent Onsager coefficients.

Now, we can consider that system $X$ is also a system of Monte Carlo harmonic oscillators\cite{art1}. In this case the Fourier's law reads (in the framework of LRT):

\begin{equation}
\lim_{t'\rightarrow t} J(t,t') = 8 \epsilon^2 \frac{E_X E_Y}{K_XK_Y}\frac{f_Xf_Y}{f_X+f_Y} (\beta_X -\beta_Y)~~.
\label{curr2osc}
\end{equation}

In a recent paper \cite{art1} we have shown that the effective
temperatures and energies of two oscillator systems coupled like in
(\ref{ham}) and evolving under a sequential dynamics never equalize. So the
heat transfer is extremely small and clearly insufficient to thermalize
degrees of freedom at different effective temperatures. The expression
(\ref{curr2osc}) justifies the results found in \cite{art1}.  Using the
asymptotic expansions for $ f_Y(t) $ and $E_Y(t) $ we can see that the
Onsager coefficient decays as $ L_{QQ} \simeq 1/(t\log^2(t)),$ and the
conductivity as $ \kappa \simeq 1/t.$ Following \cite{art1} suppose that
we couple the two oscillator systems at time $t$. If they are in the
aging state evolving at zero temperature, then they have different
effective temperatures given by $\Delta\beta=\beta_X-\beta_Y\sim
\alpha\log(t)$ with $\alpha$ a constant. The heat ${\cal Q}$ transferred
between $t$ and $t+T$ is given by,

\be
{\cal Q}=\int_t^{t+T}dxJ(x)=\alpha\int_t^{t+T}\frac{dx}{x\log(x)}
=\alpha\log\bigl(\frac{\log(t+T)}{\log(t)} \bigr)~~.
\label{eqQ}
\ee

For the two systems to equalize their temperatures the net heat transfer
between $t$ and $t+T$ must be of the order ${\cal Q}\sim
\lambda/\log(t)$ with $\lambda$ a finite constant. It can be shown that
there are two regimes for ${\cal Q}$ depending whether $T/t\ll 1$ or
$T/t\gg 1$. If $T/t\ll 1$ then ${\cal Q}\sim \alpha T/(t\log(t))$ and the
total heat transferred cannot equalize the effective temperatures. In
the other regime the heat transfer can equalize the two temperatures
only if $T\propto t$. This implies that the time
needed (in the aging regime) for compensate the difference between the
two effective temperatures is of the order of the age of the system,
then the heat transfer is unable to equalize the energies of the two
systems in a finite time-scale so the zero-th law is not valid
in the aging regime.

\section{conclusions}

We can now summarize our main results. We have established the proper
way to measure the effective temperature of a glass. It is well known
that the effective temperature based on the violation of FDT is a good
quantity in the thermodynamic sense, so in the aging regime we can
consider that the glass is in quasi-equilibrium at this effective
temperature\cite{KUR}. We have found that the correct procedure is to
couple the thermometer weakly and measure the temperature of the
thermometer which makes the heat transfer to vanish at a certain time $
t $ . In general, the measurement must be done with thermometers which
have small characteristic time-scales compared to the decorrelation time
of the aging system (in most cases this corresponds to the waiting
time).\\ To illustrate this, we have used a simple solvable model for a
glass and we have computed the heat transfer established when we couple
it with a generic thermometer. In this case the equations (\ref{beta1})
and (\ref{beta2}) were confirmed and also we have computed how the
measured temperature converges to the effective one. Note that the model
we have used is purely relaxational and only implies one large
time-scale. Now, we know that for measuring the effective temperature we
need a thermometer with a very short time-scale in comparison with the
relaxational time-scale of the system. For the case of structural
glasses is more complicated because in these systems there are two
time-scales involved. The autocorrelation function reveals a secondary
relaxation ($ \beta $-relaxation) and a primary one ($ \alpha
$-relaxation). For the secondary relaxation the system behaves as a
system which is in equilibrium and FDT holds, whereas for the primary
one, the relaxation of the system is extremely slow and FDT does not
hold. Therefore, the effective temperature we have defined describes the
violation of FDT in the long-time sector. If we are interested in the
measurement of this temperature, our results suggest that we have to
choose a thermometer with a time-scale shorter than the
$\alpha$-timescale but larger than the $\beta$-timescale in order to
avoid the effects of the initial $ \beta $-relaxation which is
associated with a different temperature from the effective
one. Otherwise, we would have a measurement where the two temperatures
involved in the relaxation could mix. Along these lines it could be very
instructive to investigate other models with two different time-scales
(for instance the Backgammon model \cite{BACK}).\\ Employing our
knowledge of the model of harmonic oscillators we have been able to
extend Fourier's law to this glassy model by considering a
time-dependent Onsager's coefficient; and we have seen how the thermal
conductivity decays. It is important to note that the decay of the
conductivity (in the case of two glasses coupled) is faster than the
decay of the energy (i.e the decay of the effective
temperature). Physically this means that, as time goes on, although the
effective temperatures of the systems are different the conductivity is
not high enough and they cannot thermalize. This seems to be the reason
for the existence of different temperatures in real glasses, in which
there are {\it fast} degrees of freedom and {\it slow} ones at different
effective temperatures which never equalize due to the smallness of the
conductivity \cite{art1}.\\ It would be very interesting to address some
other non-equilibrium features such as the long-range spatial
correlation functions for stationary non-equilibrium liquids \cite{RUBI}
for the model we have considered here and glassy systems in general.\\
{\bf Acknowledgements.} We acknowledge stimulating discussions with
S. Ciliberto and M. Mezard. We are grateful to I.Pagonabarraga and
M.Rub\'{\i} for a careful reading of the manuscript. A. G. is supported
by a grant from the University of Barcelona. F. R is supported by the
Ministerio de Educaci\'on y Ciencia in Spain (PB97-0971).

\end{document}